# Interpretation of the observations made in 1883 in Zacatecas (Mexico): A fragmented Comet that nearly hits the Earth

Hector Javier Durand Manterola[1], Maria de la Paz Ramos Lara[2] y Guadalupe Cordero[3]

[1] y [3] Departamento de Ciencias Espaciales, Instituto de Geofísica, Universidad Nacional Autónoma de México (UNAM), [2] Centro de Investigaciones Interdisciplinarias en Ciencias y Humanidades, Universidad Nacional Autónoma de México (UNAM).

hdurand_manterola@yahoo.com
ramoslm@servidor.unam.mx
gcordero@geofisica.unam.mx

Abstract

In 1883, on the 12th and 13th of August, Mexican astronomer José A. y Bonilla observed several objects passing in front of the solar disk. In 1886 in the *L'Astronomie* magazine, he reported his observations without providing a hypothesis explaining the registered phenomena. Our objective in this work is to interpret, with current knowledge, what he observed in Zacatecas. Our working hypothesis is that what Bonilla observed in 1883 was a highly fragmented comet, in an approach almost flush to the Earth's surface. The fragmentation of the comet's nucleus is a phenomenon known since the XIX century. Using the results reported by Bonilla, we can estimate the distance at which the objects approach to the Earth's surface, their size, their mass and total mass of the comet before fragmentation. According to our calculations, the distance at which the objects passed over the Earth's surface, was between 538 km and 8,062 km, the width of the objects was between 46 m and 795 m and its length between 68 m and 1,022 m, the object's mass was between $5.58 \times 10^8$ kg and $2.5 \times 10^{12}$ kg. Finally, the mass of the original comet, before fragmentation, was between $1.83 \times 10^{12}$ and $8.19 \times 10^{15}$ kg, i.e., between $2 \times 10^{-3}$ and 8.19 times the mass of Halley Comet.

1 Introduction

The Astronomical Observatory in the State of Zacatecas (Mexico) was founded on the 6$^{th}$ of December in 1882, and was under the direction of Mexican engineer and astronomer José A. y Bonilla (1853-1920), who studied astronomy in Zacatecas and in Mexico City. He studied celestial photography during his stay in the Paris Astronomical Observatory (Robles, 2010).

At that time the most important observatories in Mexico where located in the capital, the National Astronomical Observatory, under the direction of the Mexican astronomer, engineer Angel Anguiano (1840-1921), the Central Astronomical Observatory and the Central Meteorological Observatory. With the National Astronomical Observatory Mexico collaborated in the international *Carte du Ciel* project, at the end of the XIX century. The communications with the observatories in the provinces used an extensive telegraphic network (Ramos and Moreno, 2010).

The inauguration of the Zacatecas Observatory took place the same day as the rest of the observatories in Mexico were preparing to observe the transit of Venus in front of the solar disc. In Zacatecas, the observation was headed by José A. y Bonilla (from here on, only Bonilla). In the city of Puebla, it was headed by the Chief of the French commission, the French engineer Bouquet

de la Grye (1827-1909). In Guadalajara it was headed by the engineer Carlos F. Landero (head of the pacific exploration commission) and in the city of León, by José A. Brambila (Robles, 2010).

On 1$^{st}$ of January in 1886 in *L'Astronomie* magazine, which was founded by the French astronomer Camille Flammarion, Bonilla reported the passing by of a series of objects in front of the Sun on the 12$^{th}$ and 13$^{th}$ of August of 1883. These objects where surrounded by a mist and left behind a similar misty trace, such objects looked dark against the solar disc, but bright outside of this. During these days Bonilla observed 447 bodies crossing the solar disc. As the editor could not find a suitable explanation, he supposed that the objects could have been birds, insects, or dust that crossed in front of the telescope. Our objective in this paper is to interpret, using our current knowledge, what Bonilla observed in Zacatecas. Our hypothesis is that in 1883 Bonilla observed the crossing of a fragmented comet close to the Earth.

Using Bonilla's data, and known facts, we estimate: a) the distance from the Earth's surface at which the fragments crossed (between 538 km and 8,062 km). b) The width of the objects (between 46 m and 795 m). c) Its length (between 68 m and 1,022 m). d) The mass of the objects (between $5.58 \times 10^8$ kg y $2.5 \times 10^{12}$ kg). Finally, the mass of the original comet, before fragmenting, was in the $1.83 \times 10^{12}$ to $8.19 \times 10^{15}$ kg interval, i.e., between $2 \times 10^{-3}$ and 8.19 times the mass of Halley's Comet.

2 The Comet observed by Bonilla

*2.1 An estimate of the maximum distance at which the objects could have crossed.*

According to Bonilla's information (1886) the objects where not seen from the city of Puebla or Mexico City, hence for these towns the objects were not aligned with a visual towards the Sun. This allows us to calculate the maximum distance from the city of Zacatecas to the objects of 64,802 km, i.e., the real distance from the objects must have been less or at the very least, the same as this.

The previous maximum height was obtained as follows. If the objects where not seen from the City of Puebla or Mexico City, it means that these did not project on the solar disc. This indicates that, in these places, the visual to the objects together with the visual to the Sun formed an angle α greater or, at least, equal to 0.533° (median angular diameter of the Sun) (Abell, 1975) so they did not project on the Solar Disc (Figure 1).

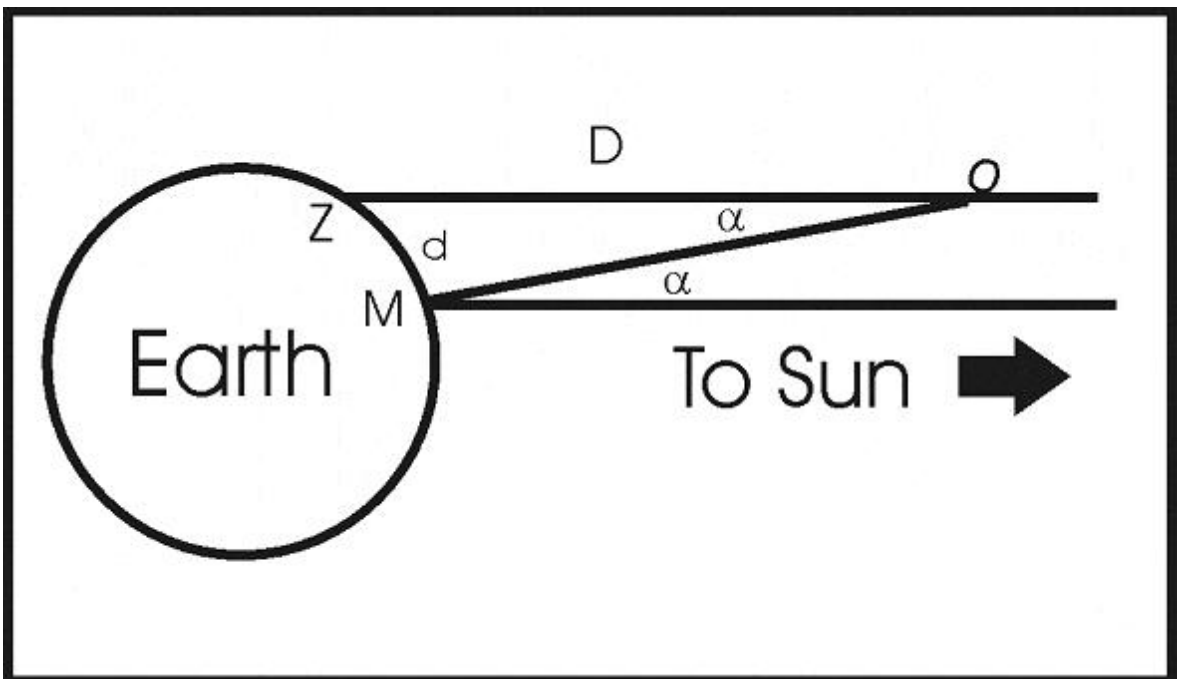

Figure 1 Geometry of Bonilla's observation: Z (Zacatecas), M (Mexico City or Puebla) and O is the point at which objects were observed. D is the distance from the Zacatecas Observatory to the objects and d is the distance between Zacatecas and Puebla or between Zacatecas and Mexico City.

The parallax of the objects with respect to the distance between Zacatecas and Puebla or Zacatecas and Mexico is the same as the angle α since they are alternate internal angles, as can be seen in figure 1.

If D >> d, the angle α in radians is

$$\alpha \approx \frac{d}{D} \qquad (1)$$

where d is the distance between Zacatecas and Puebla or Zacatecas and Mexico, and D is the distance between Zacatecas and the Objects. Solving for D we get

$$D \approx \frac{d}{\alpha} \qquad (2)$$

The mean angular diameter of the Sun $\alpha$, seen from Earth is 31' 59".3 = 0.533° = 9.305x$10^{-3}$ radians (Abell, 1975). The distance from Zacatecas to Puebla is 728 km (Internet 1). The distance from Zacatecas to Mexico City is 603 km (Internet 1). Substituting these values for d and $\alpha$ in the equation (2), the distance D to the objects is of 64,804 km, if d is the distance between Mexico and Zacatecas, and 78,238 km, if d is the distance between Puebla and Zacatecas. If the objects were at a larger distance than 78,238 km they would have been seen in Mexico City as well as in the city of Puebla. If they had been located at a distance between 64,804 km and 78,238 km they would not have been seen in Puebla but in Mexico City they would have been seen. So that both conditions of invisibility are fulfilled they must have been at a distance $\leq$ 64,804 km. Hence, the maximum distance to the objects, with the data of Bonilla, is of 64,804 km.

*2.2 Distance from Zacatecas to the objects*
If we suppose that the objects seen by Bonilla (1886) were fragments from a broken up comet, we can calculate a more accurate estimate of the distance at which the objects crossed. Our result is that the distance was less than the distance obtained in subsection 2.1 and was a distance between 538 km and 8,062 km.

The previous distance interval was obtained in the following way: If the objects moved at a velocity V and took a time t to pass through the solar disc, then the real distance X covered during their transit is (see figure 2):

$$X = Vt \qquad (3)$$

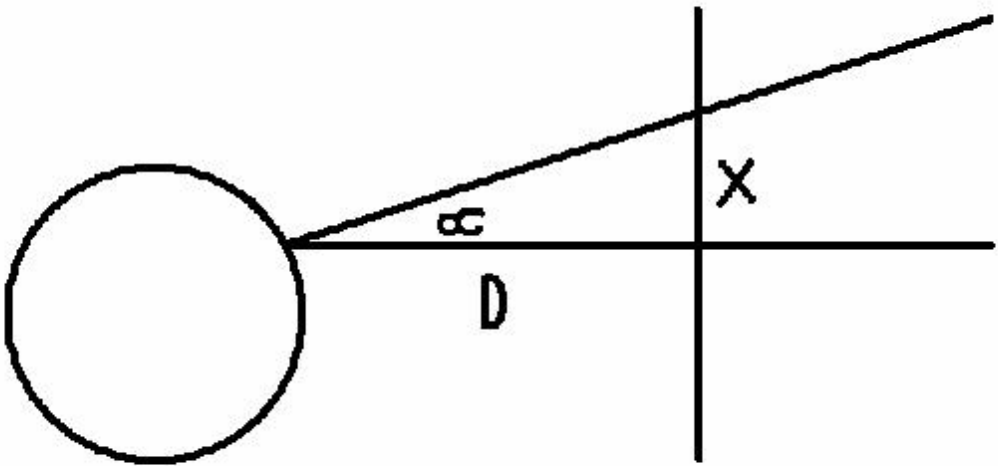

Figure 2 Geometry for estimate the distances at which objects passed

The velocity of the meteors when entering the Earth's atmosphere is between 15 and 75 km/s (Mosqueira, 1996, pp 174). We can suppose that the comets velocity when close to the Earth, is similar to these. Bonilla estimated that the duration of the objects crossing in front of the solar disc was of 1/3, ½ or 1 second. Considering these values the distance X covered in front of the Sun and estimated by the equation (3) acquires a series of values as shown in table 1.

Table 1
Estimated distances traveled crossing the solar disc

| Velocity ⇒<br>Estimated Time ⇓ | 15 km/s<br>X min ⇓ | 75 km/s<br>X max ⇓ |
|---|---|---|
| 1/3 s | 5 km | 25 km |
| ½ s | 7.5 km | 37.5 km |
| 1 s | 15 km | 75 km |

If we consider that the objects trajectory is perpendicular to the line of vision, then we get

$$\frac{X}{D} = \tan(0.533°) \tag{4}$$

Solving out D we get

$$D = \frac{X}{\tan(0.533°)} \tag{5}$$

Using equation (5) and the values of X, provided in table 1, we get the values for distance D (shown in table 2).

Table 2
Estimated distances from Zacatecas to the objects

| X (km) | D (km) |
|---|---|
| 5 | 538 |
| 7.5 | 806 |
| 15 | 1,612 |
| 25 | 2,687 |
| 37.5 | 4,031 |
| 75 | 8,062 |

8We don't know how Bonilla estimated his times since he does not mention if he had access to a chronometer to measure the seconds and fractions of a second. If these values, which Bonilla provided for the elapsed time, are correct, then the distance to the objects was between 538 km and 8,062 km.

*2.3 An estimate of the size of the objects*
We estimate that the width of the objects was between 46 and 795 m and their length between 68 and 1022 m.
The estimation of these intervals was obtained in the following way: The ratio of the size of the object (Z) to its size in the picture (L) is equal to the ratio of the length of the path in front of the Sun (X) to the size of the Sun in the picture ($\delta$). Solving out for Z we have

$$Z = \frac{L}{\delta} X \qquad (6)$$

where L is the width or the length of the objects measured on the photograph, $\delta$ the solar diameter, also measured on the photograph, and X is the real distance covered by the objects when passing through the solar disc (see § 2.2).
The width of the object in the photograph is of 0.6 to 0.7 mm. The length of the object in the photograph is of 0.9 mm and the diameter of the sun in the photograph is of 66 mm (Bonilla, 1886, editors note). With these values for L and $\delta$ and the extreme values for X (table 1) we have the values for Z in the table 3.
These values coincide with the values measured for the fragments of the 73P/Schwassmann-Wachmann 3 comet which fragmented in 2006 (Reach et al., 2009).

Table 3
Estimated sizes of the objects

|  | X ⇒ | 5 km | 75 km |
|---|---|---|---|
|  | L/δ ⇓ | ----- | ------ |
| Smaller Wide | 0.9x10-2 | 46 m | 682 m |
| Bigger Wide | 1.06x10-2 | 53 m | 795 m |
| Long | 1.36x10-2 | 68 m | 1022 m |

*2.4 An estimate of the object's mass*

The mass of every object is between $5.58 \times 10^8$ kg and $2.50 \times 10^{12}$ kg. An estimate for these values was done in the following way:

The objects were not perfect spheres (Bonilla, 1886, editor's note) and we can consider them as ellipsoid. Hence, their volume is (Internet 2)

$$V = \frac{4}{3}\pi abc \qquad (7)$$

where a, b and c are the three semi axes of the ellipsoid.

Since we can only obtain two semi axes from the measurements of the photograph, then we can suppose that the third axis is the same as one of the ones measured. Hence, their volume will be

$$V = \frac{4}{3}\pi ab^2 \qquad (8)$$

The bodies' mass will be:

$$M = \rho V = \frac{4}{3}\rho\pi ab^2 \qquad (9)$$

where ρ is the density.

If we suppose that the objects where made of ice, then ρ = 925 kg/m$^3$. Using the equation (9) and the estimated sizes in section 2.3, then we get an estimate of the mass of the bodies (table 4).

Table 4
Estimated Mass of the objects in kg

| b (m) → | 46 | 53 | 682 | 795 |
|---|---|---|---|---|
| a (m) ↓ | | | | |
| 68 | 5.58E+08 | 7.40E+08 | 1.23E+11 | 1.67E+11 |
| 1022 | 8.38E+09 | 1.11E+10 | 1.84E+12 | 2.50E+12 |

*2.5 Estimation of the original comet's mass, before fragmenting*

We estimate that the original mass of the comet was between $1.83 \times 10^{12}$ and $8.19 \times 10^{15}$ kg, i.e., $2 \times 10^{-3}$ to 8.19 times the mass of Halley's Comet.

This estimation was made as follows:

The total time for Bonilla's observation was 3 h and 25 min between both days and he observed a total of 447 objects (Bonilla, 1886). This implies an average of 131 objects per hour. So, from 8 o'clock in the morning on the 12$^{th}$ to 9 o'clock in the morning of the 13$^{th}$ should have passed 3275 objects. If we take the masses provided in section 2.4 and multiply them by this number, then the comet's original mass was between $1.83 \times 10^{12}$ and $8.19 \times 10^{15}$ kg. The estimated mass for Halley's Comet is between $10^{14}$ and $10^{15}$ kg (Binzel et al., 2000, pp 322), so the mass for Bonilla's Comet was between $2 \times 10^{-3}$ and 8.19 times the mass of Halley's Comet.

3 Discussion

The only bodies in the Solar System which are surrounded by a bright mistiness are the comets, so it is appropriate to suppose that the objects seen by Bonilla were small comets. The integrated mass of all the objects is of the same order of the masses known for cometary nucleus; this supports the idea that what Bonilla saw was a comet.

The fragmenting of the cometary nucleus is a phenomenon known since the XIX century, for example, the Biela comet, which in 1845 split into two fragments (de Pater and Lissauer, 2001). In the last 150 years we have observed more than 40 comets which have experienced fragmentation (Boehnhardt, 2004; Fuse et al., 2007; Reach et al., 2009). But during Bonilla's time only two fragmentations had been observed, 3D/Biela comet (fragmented into two pieces) and C1860 D1 Liais comet (Boehnhardt 2004). Neither of these two comets had been fragmented so much as the comet observed by Bonilla. It was probably for this reason that neither Bonilla nor the editor of *L'Astronomie* thought in this explanation.

Fragments of comets with a size as small as what we calculated, would loose their volatiles in a very short time, then we suppose that the fragmenting of the major object happened a little before Bonilla's observation.

It is well known that during the 12th and 13th of August of every year a shower of shooting stars known as Perseids can be observed. What one can assume is that Bonilla saw these bodies, although, the Perseids are seen throughout the entire northern hemisphere so we can assume that the trajectory path for Perseids is wider than the Earth. On the other hand with Bonilla's observation trajectory path for these objects looks much collimated (This was not seen in Mexico or Puebla).

During 1883 two comets were seen: 1883 I (Brooks-Swift) and 1883 II (Pons-Brooks) (Cincinnati Observatory, 1885). The fragmenting of any of the two comets some months or years before 1883 could explain Bonilla's objects. In one of Bonilla's figures, Bonilla shows that the trajectories are very slanted in a northeast to south west direction (~43° measured from the North) (Bonilla, 1886, figure 118). On the other hand, the slanting of the orbit of the Pons-Brooks comet with respect to the ecliptic was of 74° (Cincinnati Observatory, 1885), which would mean an angle between 7° and 39° with respect to the North. If we take the greater angle, this would coincide with Bonilla's observations.

It also could have happened that the objects that Bonilla observed were pieces of a comet different to the other two seen this year and which pieces were not bright enough. A similar, but more recent case, which can illustrate what happened in 1883, is the case of the Schwassmann-Wachmann 3 (Fuse et al., 2007; Reach et al., 2009), fragmented in its three passes of 1995/1996, 2001 and 2006. The great fragmentation of this comet during 2006 can be seen in the photograph in figure 3.

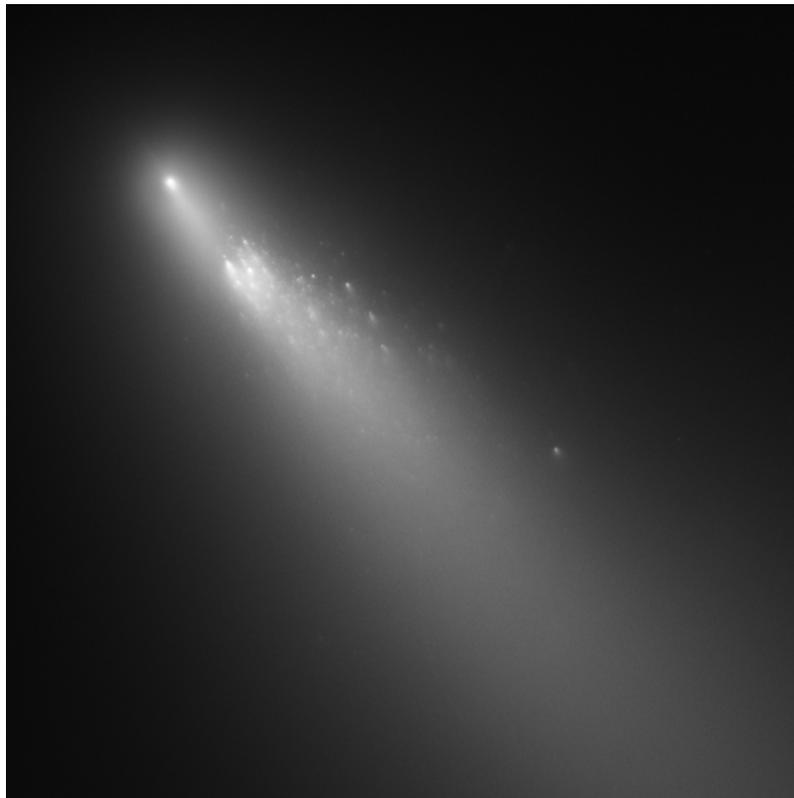

Figure 3 Schwassmann-Wachmann3 Comet: Fragment B photographed during its passing in 2006. We can observe up to 73 fragments. This photo is from APOD (Astronomy Picture of the Day) (Internet 3). Credit: NASA, ESA, H. Weaver (JHU/APL), M. Mutchler and Z. Levay (STScI)

Why, the objects seen by Bonilla, were not seen from any other place? With respect to Mexico City and Puebla it has already been explained that in these places the objects was not projected on the solar disc due to their closeness to Earth. But at the latitude of Zacatecas, it seems, that there was not any observatory. Since the parallel at this latitude goes through the Atlantic, the Sahara Desert, Arabia, North of India, Southeast Asia and the Pacific Ocean. And since the passing was on the day side any of the bodies that could have entered the high atmosphere would not have been seen. Also, if Bonilla's comet was telescopic, as is the case of the Schwassmann-Wachmann 3 comet, it would not have been seen easily.

For the calculated distances, we see that these objects were close to impact Earth. Furthermore, the calculated size of the objects is greater than or of the same order of the object which produced the Tunguska event. So if they had collided with Earth we would have had 3275 Tunguska events in two days, probably an extinction event.